\documentclass[pra,twocolumn,superscriptaddress,amsmath,amssymb]{revtex4}
\usepackage{epsfig}
\usepackage{subfigure}
\usepackage{amssymb,amsmath}
\usepackage{graphicx}
\usepackage{epstopdf}
\usepackage{epsfig}
\usepackage{color}
\usepackage{amsfonts}
\usepackage{amscd}
\usepackage{hyperref}
\usepackage{bbm}
\usepackage{mathrsfs}

\DeclareMathSizes{10}{10}{7}{5}

\begin{document}
\title{Non-Markovian quantum state diffusion for an open quantum system in fermionic environments}
\date{\today}
\author{Mi Chen}
\affiliation{Department of Physics and State Key Laboratory of
Surface Physics, Fudan University, Shanghai 200433, China}
\author{J. Q. You}
\affiliation{Beijing Computational Science Research Center, Beijing
100084, China}

\begin{abstract}
Non-Markovian quantum state diffusion (NMQSD) provides a powerful
approach to the dynamics of an open quantum system in bosonic
environments. Here we develop an NMQSD method to study the open
quantum system in fermionic environments. This
problem involves anticommutative noise functions (i.e., Grassmann
variables) that are intrinsically different from the noise functions
of bosonic baths. We obtain the NMQSD equation for quantum states of
the system and the non-Markovian master equation. Moreover, we apply
this NMQSD method to single and double quantum-dot systems.
\end{abstract}
\pacs{03.65.Yz, 05.40.-a, 73.21.La}
\keywords{}
\maketitle

\section{Introduction}

The theory of open quantum systems has
become an increasingly important topic in, e.g., quantum information
science, quantum measurement, and quantum optics. Traditionally, the
dynamics of an open quantum system was often investigated using a
Markov master equation derived by invoking the Born-Markov
appoximation. However, this formalism fails for many solid-state
systems (see, e.g., \cite{jqyou1}) where the system-environment
coupling is strong and the environment is structured. Thus a
non-Markovian master equation is required when considering the
memory effect and back action of the environment. It is known that
the derivation of an exact non-Markovian master equation has long
been a challenging task. One of the breakthroughs is the exact
non-Markovian master equation for quantum Brownian motion model derived
by Hu {\it et al.}~\cite{Hu} using the Feynman-Vernon influence
functional path-integral method~\cite{Feynman}.

Of all the theoretical strategies used to deal with open quantum
systems, a non-Markovian quantum trajectory theory known as
non-Markovian quantum state diffusion (NMQSD)~\cite{Diosi,Strunz1}
provides a powerful approach to the dynamics of an open quantum
system in bosonic environments. In this approach, when the so-called
$O$-operator is obtained, the quantum dynamics of an open system is
determined by solving the NMQSD equation (i.e., a diffusive
stochastic Schr\"{o}dinger equation) and the non-Markovian master
equation can also be derived~\cite{Yu1,Strunz2}. In contrast to the
conventional master equation under the Born approximation, this
non-Markovian mater equation is derived non-perturbatively, so it
applies even for a strong system-environment coupling. Indeed, some
exact $O$-operators have been found in a variety of quantum
models~\cite{Diosi,Strunz1,Yu1,Strunz2,Jing1,Eberly}, including
multilevel models~\cite{Jing2}.

In addition to bosonic baths, fermionic baths are also involved in
many physical systems, particularly in solid-state systems. The
Feynman-Vernon influence functional path-integral method can also be
used to study the quantum dynamics of an open system in a fermionic
environment (see, e.g., \cite{Hedegard}). Recently, this
path-integral method was extended to derive non-Markovian master
equations for nanodevices~\cite{Zhang1,Zhang2}. Also, there were
other studies on quantum dynamics and transport of nanostructured
systems, including the rate-equation approach~\cite{Gurvitz,Stoof},
and the Markovian (see, e.g., \cite{Haug1,Cottet}) and non-Markovian
(see, e.g., \cite{Goan}) master equation approaches. In addition,
the nonequilibrium Green's function method can be used to study the
quantum transport through the nanostructures (see,
e.g.,~\cite{Haug2}). Nevertheless, the extension of NMQSD method to
an open quantum system in fermionic baths has been a long-standing
unsolved problem because this open system involves anticommutative
noise functions (i.e., Grassmann variables) that are intrinsically
different from the noise functions of bosonic baths.

In this paper,
we develop an NMQSD method to study the open quantum system in
fermionic baths. This NMQSD approach is formulated in a
non-perturbative manner and it applies for both weak and strong
system-environment couplings. We not only obtain the NMQSD equation
for quantum states of the system, but also derive the non-Markovian
master equation. Moreover, as interesting examples, we apply this
NMQSD method to single and double quantum-dot systems.
Note that our NMQSD method belongs to the quantum trajectory approach involving continuous time evolution. There is another kind of quantum trajectory approach which involves discontinuous time evolution, i.e., quantum jumps (see, e.g., \cite{Dali}). The non-Markovian quantum trajectory approach in \cite{Piilo} generalizes the Markovian quantum jump method and can also apply to both bosonic and fermionic baths.

\section{Quantum state diffusion equation}

We consider a quantum
system coupled to two fermionic baths: $H=H_{\rm sys}+H_{\rm
env}+H_{\rm int}$, with (we set $\hbar=1$)
\begin{align}
&H_{\rm env}= \sum_{k}(\omega_{Lk}a_{Lk}^\dagger a_{Lk}+
\omega_{Rk}a_{Rk}^\dagger a_{Rk}), \label{bath} \\
&H_{\rm int}=
\sum_{k}(g_{Lk}c_{L}^\dagger a_{Lk}+ g_{Rk}c_{R}^\dagger a_{Rk}+
{\rm H.c.}). \label{int}
\end{align}
Here $H_{\rm sys}$ denotes the Hamiltonian of the system, $H_{\rm
env}$ is the Hamiltonian of the two electric leads acting as
fermionic baths, and $H_{\rm int}$ models the interactions between
the system and the two baths. The spectral density function of each
bath is
\begin{equation}
J_{\lambda}(\omega)=\sum_{k}\lvert g_{\lambda
k}\rvert^2\delta(\omega-\omega_{\lambda k}),
\end{equation}
where $\lambda=L$ or
$R$. In Eq.~(\ref{bath}), $a_{\lambda k}^\dagger$ ($a_{\lambda k}$)
is the fermionic creation (annihilation) operator for a quantum
state with wave vector $k$ in the left or right lead. We assume that
the system of interest couples to the two leads via single channels
characterized by the fermionic creation (annihilation) operators
$c_{\lambda k}^\dagger$ ($c_{\lambda k}$) [see Eq.~(\ref{int})].
Extension to a multi-channel case is straightforward.

In an NMQSD approach, environments are required to be initially at
zero temperature, so as to conveniently represent the environmental
degrees of freedom with the coherent state basis. As for
environments initially with a nonzero temperature, one can map the
nonzero-temperature density operator to a zero-temperature density
operator using a Bogoliubov transformation~\cite{Yu2}. In the case
of fermionic baths, this requires to add $\sum_{\lambda
k}\omega_{\lambda k}b_{\lambda k} b_{\lambda k}^\dagger$ to
Eq.~(\ref{bath}), corresponding to the part involving holes in the
electric leads. The Bogoliubov transformation for fermionic
operators can be introduced as
\begin{align}
& a_{\lambda k}= \sqrt{1- \bar{n}_{\lambda k}}
d_{\lambda k}- \sqrt{\bar{n}_{\lambda k}}e_{\lambda k}^\dagger \nonumber\\
& b_{\lambda k}= \sqrt{1- \bar{n}_{\lambda k}}e_{\lambda k}+
\sqrt{\bar{n}_{\lambda k}}d_{\lambda k}^\dagger, \label{Bog}
\end{align}
where $\bar{n}_{\lambda k}=[e^{(\omega_{\lambda
k}-\mu_{\lambda})/k_{B}T}+1]^{-1}$ is the average number of
electrons in the $k$th state of the left (right) electric lead with
chemical potential $\mu_{\lambda}$. In Eq.~(\ref{Bog}), the
coefficients are determined by the requirement that the derived
master equation reduces to a Lindblad form in the Markovian limit.
The transformed Hamiltonian $\mathcal{H}$ is written as
\begin{eqnarray}
\mathcal{H}\!&\!=\!&\! H_{\rm sys}+\sum_{\lambda
k}\left[\omega_{\lambda k}(d_{\lambda k}^\dagger d_{\lambda
k}+e_{\lambda k}e_{\lambda
k}^\dagger)\right.\nonumber\\
&\!&\!+\left. (\sqrt{\bar{n}_{\lambda k}} g_{\lambda k}^*
c_{\lambda}e_{\lambda k}+\sqrt{1-\bar{n}_{\lambda k}} g_{\lambda
k}c_{\lambda}^\dagger d_{\lambda k}+ {\rm H.c.})\right],~~~~
\end{eqnarray}
where the new fermionic operator $d_{\lambda k}$ ($e_{\lambda
k}^\dagger$) corresponds to the annihilation of electrons (holes) in
the virtual fermionic baths. Note that the effects of temperature
are incorporated into the transformed Hamiltonian and the fermionic
baths with {\it nonzero} initial temperatures are mapped to virtual
fermionic baths with {\it zero} initial temperature.

In the interaction picture with respect to the environmental
Hamiltonian $\mathcal{H}_{\rm env}= \sum_{\lambda k} \omega_{\lambda
k}(d_{\lambda k}^\dagger d_{\lambda k}+e_{\lambda k}e_{\lambda
k}^\dagger)$, the total Hamiltonian reads
\begin{eqnarray}
\mathcal{H}(t)\!&\!=\!&\! H_{\rm sys}+ \sum_{\lambda
k}(\sqrt{\bar{n}_{\lambda k}} g_{\lambda k}^* c_{\lambda}e_{\lambda k}e^{i\omega_{\lambda k}t}\nonumber\\
&&\!+\sqrt{1- \bar{n}_{\lambda k}} g_{\lambda k}c_{\lambda}^\dagger
d_{\lambda k}e^{-i\omega_{\lambda k}t}+ {\rm H.c.}),
\end{eqnarray}
and the quantum state of the total system satisfies the equation of
motion
\begin{equation}
\partial_{t}\lvert\Psi_{t}\rangle=
-i\mathcal{H}(t)\lvert\Psi_{t}\rangle. \label{EOM}
\end{equation}
We assume that the quantum state of the total system is factorized
at the initial time $t=0$, so that $\lvert\Psi_{0}\rangle=\lvert
\varphi_{0}\rangle\otimes\lvert 0\rangle$, with the virtual
fermionic baths initially in the ground state (i.e., at zero
temperature): $\lvert 0\rangle=\bigotimes_{\lambda}\lvert
0\rangle_{\lambda d}\otimes\lvert 0\rangle_{\lambda e}$, where
$d_{\lambda k}\lvert 0\rangle=0$, and $e_{\lambda k}\lvert
0\rangle=0$.

Define a femionic coherent-state basis for the environmental degrees
of freedom:
\begin{equation}
\lvert zw\rangle=\bigotimes_{\lambda}\lvert
z\rangle_{\lambda}\otimes\lvert w\rangle_{\lambda},
\end{equation}
with
\begin{eqnarray}
\lvert z\rangle_{\lambda}\!&\!=\!&\! \bigotimes_{k}\lvert
z_{k}\rangle_{\lambda}
= e^{-\sum_{k}z_{\lambda k}d_{\lambda k}^\dagger}\lvert 0\rangle \nonumber\\
\lvert w\rangle_{\lambda}\!&\!=\!&\! \bigotimes_{k}\lvert
w_{k}\rangle_{\lambda}= e^{-\sum_{k}w_{\lambda k}e_{\lambda
k}^\dagger}\lvert 0\rangle,
\end{eqnarray}
where $z_{k}$ and $w_{k}$ are Grassmann variables that obey the
anticommutation relation. With the completeness relation for
coherent states
$\int e^{-z^*z-w^*w}\lvert zw \rangle\langle zw \rvert d^2 z d^2
w=1$,
the state $\lvert\Psi_{t}\rangle$ can be expressed as
\begin{equation}
\label{state2}\lvert\Psi_{t}\rangle= \int e^{-z^*z-w^*w}\lvert
zw\rangle\otimes\lvert \psi_{t}(z^*,w^*)\rangle d^2 z d^2 w,
\end{equation}
where
\begin{eqnarray}
&&z^*z\equiv\sum_{\lambda k}z^*_{\lambda k}z_{\lambda k},
~~~w^*w\equiv\sum_{\lambda k}w^*_{\lambda k}w_{\lambda k},\nonumber\\
&&d^2z\equiv\prod_{\lambda k}dz_{\lambda k}^*dz_{\lambda k},
~~~d^2w\equiv\prod_{\lambda k}dw_{\lambda k}^*dw_{\lambda k}.~~~
\end{eqnarray}
The
actions of annihilation (creation) operators $d_{\lambda k}$ and
$e_{\lambda k}$ ($d_{\lambda k}^{\dagger}$ and $e_{\lambda
k}^{\dagger}$) on fermionic coherent states satisfy the
relations~\cite{Atland}:
\begin{eqnarray}
&&d_{\lambda
k}|z\rangle_{\lambda}=z_{\lambda k}|z\rangle_{\lambda},~~~ d_{\lambda
k}^{\dagger}|z\rangle_{\lambda}=-\frac{\partial}{\partial z_{\lambda
k}}|z\rangle_{\lambda},\nonumber\\
&&e_{\lambda k}|w\rangle_{\lambda}=w_{\lambda k}|w\rangle_{\lambda},
~~~e_{\lambda k}^{\dagger}|w\rangle_{\lambda}=-\frac{\partial}{\partial w_{\lambda
k}}|w\rangle_{\lambda}.~~~~~~
\end{eqnarray}
When projecting onto the coherent-state
basis, the equation of motion (\ref{EOM}) can be reduced to the
NMQSD equation for a pure state of the system
$\lvert\psi_{t}(z^*,w^*)\rangle\equiv \langle
zw\rvert\Psi_{t}\rangle$:
\begin{eqnarray}
\frac{\partial}{\partial t}\lvert\psi_{t}\rangle\!&\!=\!&\!-iH_{\rm
sys}\lvert\psi_{t}\rangle
-\sum_{\lambda}\left[c_{\lambda}z_{\lambda}^*(t)\lvert\psi_{t}\rangle
+c_{\lambda}^\dagger w_{\lambda}^*(t)\lvert\psi_{t}\rangle\right]
\nonumber\\
&&\!-\sum_{\lambda}c_{\lambda}^\dagger\int_{0}^t\alpha_{\lambda
1}(t-s) \frac{\delta}{\delta
z_{\lambda}^*(s)}\lvert\psi_{t}\rangle ds \nonumber\\
&&\!-\sum_{\lambda}c_{\lambda}\int_{0}^t\alpha_{\lambda 2}(t-s)
\frac{\delta}{\delta w_{\lambda}^*(s)}\lvert\psi_{t}\rangle ds,
\label{QSD-1}
\end{eqnarray}
which initiates from $\lvert\psi_{t=0}(z^*,w^*)\rangle=\lvert
\varphi_{0}\rangle$. Here the noise functions $z_{\lambda}^*(t)$ and
$w_{\lambda}^*(t)$ are defined as
\begin{align}
&z_{\lambda}^*(t)=
-i\sum_{k}\sqrt{1-\bar{n}_{\lambda k}}
g_{\lambda k}^*z_{\lambda k}^*e^{i\omega_{\lambda k}t}, \nonumber\\
&w_{\lambda}^*(t)= -i\sum_{k}\sqrt{\bar{n}_{\lambda k}}g_{\lambda k}w_{\lambda
k}^*e^{-i\omega_{\lambda k}t}.
\end{align}
The temperature-dependent environment correlation functions are
\begin{eqnarray}
\alpha_{\lambda1}(t-s)\!&\!\equiv\!&\!\mathcal{M}\{z_{\lambda}(t)z_{\lambda}^*(s)\}\nonumber\\
\!&\!=\!&\!\int
d\omega[1-\bar{n}_{\lambda}(\omega)]J_{\lambda}(\omega)e^{-i\omega(t-s)},\nonumber\\
\alpha_{\lambda2}(t-s)\!&\!\equiv\!&\!\mathcal{M}\{w_{\lambda}(t)w_{\lambda}^*(s)\}\nonumber\\
\!&\!=\!&\!\int
d\omega\bar{n}_{\lambda}(\omega)J_{\lambda}(\omega)e^{i\omega(t-s)},
\end{eqnarray}
where $\mathcal{M}\{\cdot\}$ denotes the statistical mean over all
noise variables: $\mathcal{M}\{\cdot\}\equiv\int
e^{-z^*z-w^*w}\{\cdot\} d^2z d^2w$.

Introducing $O$-operators by
\begin{eqnarray}
\frac{\delta}{\delta
z_{\lambda}^*(s)}\lvert\psi_{t}(z^*,w^*)\rangle\!&\!=\!&\!O_{\lambda
1}(t,s,z^*,w^*)\lvert\psi_{t}(z^*,w^*)\rangle, \nonumber\\
\frac{\delta}{\delta
w_{\lambda}^*(s)}\lvert\psi_{t}(z^*,w^*)\rangle\!&\!=\!&\!O_{\lambda
2}(t,s,z^*,w^*)\lvert\psi_{t}(z^*,w^*)\rangle,~~~
\end{eqnarray}
we can write the
NMQSD equation in a time-local form:
\begin{eqnarray}
\frac{\partial}{\partial t}\lvert\psi_{t}\rangle\!&\!=\!&\!-iH_{\rm
sys}\lvert\psi_{t}\rangle-\sum_{\lambda}
\left[c_{\lambda}z_{\lambda}^*(t)+c_{\lambda}^\dagger
w_{\lambda}^*(t)\right.\nonumber\\
&&\!\left.+c_{\lambda}^\dagger\bar{O}_{\lambda
1}(t,z^*,w^*)+c_{\lambda}\bar{O}_{\lambda
2}(t,z^*,w^*)\right]\lvert\psi_{t}\rangle,~~~~~~\label{QSD-2}
\end{eqnarray}
where $\bar{O}_{\lambda n}\equiv\int_{0}^tds\alpha_{\lambda
n}(t-s)O_{\lambda n}(t,s,z^*,w^*)$, $n=1,2$.

With the consistency conditions
\begin{equation}
\frac{\partial}{\partial t}
\frac{\delta\lvert\psi_{t}\rangle}{\delta z_{\lambda}^*(s)}
=\frac{\delta}{\delta z_{\lambda}^*(s)}
\frac{\partial\lvert\psi_{t}\rangle}{\partial t},~~~
\frac{\partial}{\partial
t}\frac{\delta\lvert\psi_{t}\rangle}{\delta w_{\lambda}^*(s)}
=\frac{\delta}{\delta
w_{\lambda}^*(s)}\frac{\partial\lvert\psi_{t}\rangle}{\partial t},
\end{equation}
as well as the initial conditions $O_{\lambda
1}(t,s,z^*,w^*)\lvert_{t=s}=c_{\lambda}$ and $O_{\lambda
2}(t,s,z^*,w^*)\lvert_{t=s}= c_{\lambda}^\dagger$, we obtain the
equations of motion for the $O$-operators:
\begin{eqnarray}
\frac{\partial O_{\lambda n}}{\partial t}\!&\!=\!&\! [-iH_{\rm
sys}-\sum_{\lambda'}(c_{\lambda'}^\dagger\bar{O}_{\lambda' 1}
+c_{\lambda'}\bar{O}_{\lambda' 2}), O_{\lambda n}]+Q_n\nonumber\\
&&\!+\sum_{\lambda'}\left(\{c_{\lambda'},O_{\lambda n}\}
z_{\lambda'}^*(t)+\{c_{\lambda'}^\dagger,O_{\lambda n}\}
w_{\lambda'}^*(t)\right),\nonumber\\ \label{EOM-O}
\end{eqnarray}
where the square and curly brackets denote the commutator and
anticommutator, respectively, and
\begin{equation}
Q_n=c_{L}^\dagger\frac{\delta\bar{O}_{L
1}}{\delta\Lambda_n}+c_{R}^\dagger\frac{\delta\bar{O}_{R
1}}{\delta\Lambda_n}+c_{L}\frac{\delta\bar{O}_{L
2}}{\delta\Lambda_n}+c_{R}\frac{\delta\bar{O}_{R
2}}{\delta\Lambda_n},
\end{equation}
with $\Lambda_1=z_{\lambda}^*(s)$, and $\Lambda_2=w_{\lambda}^*(s)$.

\section{Master equation}

The reduced density operator of an
open quantum system by tracing over the environmental degrees of
freedom can be obtained by taking the statistical mean for a density
operator related to the state $\lvert\psi_{t}(z^*,w^*)\rangle$:
\begin{equation}
\rho_{t}={\rm Tr}_{\rm
env}\lvert\Psi_{t}\rangle\langle\Psi_{t}\rvert
=\mathcal{M}\{P_t\},
\end{equation}
where
$P_t\equiv\lvert\psi_{t}(z^*,w^*)\rangle\langle\psi_{t}(-z,-w)\rvert$.

Using the relation
\begin{eqnarray}
\frac{\partial P_{t}}{\partial t}\!&\!=\!&\!\frac{\partial
\lvert\psi_{t}(z^*,w^*)\rangle}{\partial
t}\langle\psi_{t}(-z,-w)\rvert \nonumber\\
&&\!+
\lvert\psi_{t}(z^*,w^*)\rangle\frac{\partial\langle\psi_{t}(-z,-w)\rvert}{\partial
t},
\end{eqnarray}
and Eq.~(\ref{QSD-2}), we derive the following non-Markovian
master equation:
\begin{eqnarray}
\frac{\partial\rho_{t}}{\partial t}\!&\!=\!&\! -i[H_{\rm
sys},\rho_{t}]+
\sum_{\lambda}\left([c_{\lambda},\mathcal{M}\{P_{t}\bar{O}_{\lambda 1}^\dagger(t,-z,-w)\}] \right.\nonumber\\
&&\!-[c_{\lambda}^\dagger,\mathcal{M}\{\bar{O}_{\lambda 1}(t,z^*,w^*)P_{t}\}]\nonumber\\
&&\!-[c_{\lambda},\mathcal{M}\{\bar{O}_{\lambda
2}(t,z^*,w^*)P_{t}\}] \nonumber\\
&&\!\left.+[c_{\lambda}^\dagger,\mathcal{M}\{P_{t}\bar{O}_{\lambda
2}^\dagger(t,-z,-w)\}]\right). \label{g-master}
\end{eqnarray}
This master equation is derived non-perturbatively, so it applies
even for a strong coupling between the system and the environments.
Moreover, in addition to trace preserving, it also preserves the
positivity and hermiticity.

In the Markovian limit, there are
\begin{eqnarray}
\alpha_{\lambda
1}(t-s)\!&\!\rightarrow\!&\!(1-\bar{n}_{\lambda})\Gamma_{\lambda}\delta(t-s),\nonumber\\
\alpha_{\lambda 2}(t-s)\!&\!\rightarrow\!&\!\bar{n}_{\lambda}\Gamma_{\lambda}\delta(t-s),
\end{eqnarray}
where $\Gamma_{\lambda}= 2\pi\rho_{\lambda}\lvert
g_{\lambda}\rvert^2$, with $\lambda=L$ ($R$), is the electron
tunneling rate between the system and the left (right) lead. Also,
the time-integrated $O$-operators become
\begin{eqnarray}
\bar{O}_{\lambda 1}\!&\!\rightarrow\!&\!\frac{1}{2}\Gamma_{\lambda}(1-\bar{n}_{\lambda})c_{\lambda},\nonumber\\
\bar{O}_{\lambda 2}\!&\!\rightarrow\!&\!
\frac{1}{2}\Gamma_{\lambda}\bar{n}_{\lambda}c_{\lambda}^\dagger.
\end{eqnarray}
Therefore, the master equation (\ref{g-master}) is reduced to
\begin{eqnarray}
\frac{\partial\rho_{t}}{\partial t}\!&\!=\!&\! -i[H_{\rm sys},\rho_{t}]\nonumber\\
&&\!+\sum_{\lambda}\frac{\Gamma_{\lambda}}{2}\left[\bar{n}_{\lambda}
(2c_{\lambda}^\dagger \rho_{t}c_{\lambda}
- c_{\lambda}c_{\lambda}^\dagger\rho_{t}- \rho_{t}c_{\lambda}c_{\lambda}^\dagger)\right.\nonumber\\
&&\!+\left.(1-\overline{n}_{\lambda})
(2c_{\lambda}\rho_{t}c_{\lambda}^\dagger- c_{\lambda}^\dagger
c_{\lambda}\rho_{t}- \rho_{t}c_{\lambda}^\dagger
c_{\lambda})\right].~~~ \label{m-master}
\end{eqnarray}
It is clear that this Markov master equation has a Lindblad form.

\section{Application to quantum-dot systems}

Below we apply our NMQSD approach to single and double quantun-dot systems.

\subsection{Single quantum dot}

Suppose that the single quantum dot is
in the strong Coulomb blockade regime, so that only one electron is
allowed therein. The Hamiltonian of the system is written as
\begin{equation}
H_{\rm sys}=\omega_{0}c^\dagger c,
\end{equation}
and $c_{L}=c_{R}=c$ for $H_{\rm int}$ in Eq. (\ref{int}).
%
The non-Markovian master equation is exactly derived as (see Appendix A)
\begin{align}
\frac{\partial\rho_{t}}{\partial t}=&-i[H_{\rm sys},\rho_{t}]
+\Gamma_{1}(t)[c,\rho_{t}c^\dagger]+
\Gamma_{2}(t)[c,c^\dagger\rho_{t}]\nonumber\\
&+ \Gamma_{1}^*(t)[c\rho_{t},c^\dagger]+
\Gamma_{2}^*(t)[\rho_{t}c,c^\dagger], \label{dot-master}
\end{align}
with time-dependent rates
\begin{equation}
\Gamma_{j}(t)=\int_{0}^t[\alpha_{1}(s-t)A_j(t,s)-\alpha_{2}(t-s)B_j(t,s)]ds,
\label{coefficient a}
\end{equation}
where $\alpha_{j}(t)=\alpha_{Lj}(t)+\alpha_{Rj}(t)$; $A_j(t,s)$ and
$B_j(t,s)$ are determined by the integro-differential equations:
\begin{align}
&(\frac{\partial}{\partial s}-i\omega_{0})A_j(t,s)+
\int_{0}^s\beta(s-s')A_j(t,s')ds'= U(t,s)\nonumber\\
&(\frac{\partial}{\partial s}-i\omega_{0})B_j(t,s)+
\int_{0}^s\beta(s-s')B_j(t,s')ds'= V(t,s),
\end{align}
with
\begin{eqnarray}
&&\beta(s-s')\equiv\alpha_{1}(s'-s)+\alpha_{2}(s-s'),\nonumber\\
&&U(t,s)\equiv\int_{0}^t\alpha_{2}(s-s')h(t,s')ds',\nonumber\\
&&V(t,s)\equiv\int_{0}^t\alpha_{1}(s'-s)h(t,s')ds',
\end{eqnarray}
and the final
conditions at $s=t$: $A_1(t,t)=B_1(t,t)=1$, and
$A_2(t,t)=B_2(t,t)=0$. Here $h(t,s)$ satisfies the equation
\begin{equation}
(\frac{\partial}{\partial s}-i\omega_{0})h(t,s)-
\int_{s}^t\beta(s-s')h(t,s')ds'= 0,
\end{equation}
with the final condition $h(t,t)=1$.

In the Markovian limit, there are
\begin{eqnarray}
\Gamma_{1}(t)\!&\!\rightarrow\!&\!\frac{1}{2}[1-\bar{n}_{L}(\omega_{0})]\Gamma_{L}
+\frac{1}{2}[1-\bar{n}_{R}(\omega_{0})]\Gamma_{R}, \nonumber\\
\Gamma_{2}(t)\!&\!\rightarrow\!&\!-\frac{1}{2}\bar{n}_{L}(\omega_{0})\Gamma_{L}
-\frac{1}{2}\bar{n}_{R}(\omega_{0})\Gamma_{R}.
\end{eqnarray}
Let us consider the
zero-temperature case with $\bar{n}_{\lambda}(\omega_{0})\rightarrow
\theta(\mu_{\lambda}- \omega_{0})$, were $\theta$ is the Heaviside
step function. If the single-dot level $\omega_0$ lies within the
energy window $\mu_{L}>\omega_{0}>\mu_{R}$, the master equation
(\ref{dot-master}) reduces to
\begin{eqnarray}
\frac{\partial}{\partial t}\rho_{t}\!&\!=\!&\!-i\omega_{0}[c^\dagger
c,\rho_{t}]+ \frac{1}{2}\Gamma_L(2c^\dagger\rho_{t}c-
cc^\dagger\rho_{t}- \rho_{t}cc^\dagger)\nonumber\\
&&\!+\frac{1}{2}\Gamma_R(2c\rho_{t}c^\dagger- c^\dagger c\rho_{t}-
\rho_{t}c^\dagger c).
\end{eqnarray}
With the basis state $|0\rangle$ ($|1\rangle$) which denotes an
empty (occupied) dot, it follows that the density matrix elements
satisfy
\begin{eqnarray}
&&\dot{\rho}_{00}= -\Gamma_{L}\rho_{00}+ \Gamma_{R}\rho_{11},\nonumber\\
&&\dot{\rho}_{11}= \Gamma_{L}\rho_{00}- \Gamma_{R}\rho_{11},\nonumber\\
&&\dot{\rho}_{10}= -(i\omega_{0}+ \Gamma_{L}+ \Gamma_{R})\rho_{10},
\end{eqnarray}
which are exactly the rate equations obtained by Gurvitz and
Prager~\cite{Gurvitz}.

\begin{figure}
\includegraphics[width=3.4in,
bbllx=52,bblly=335,bburx=536,bbury=619]{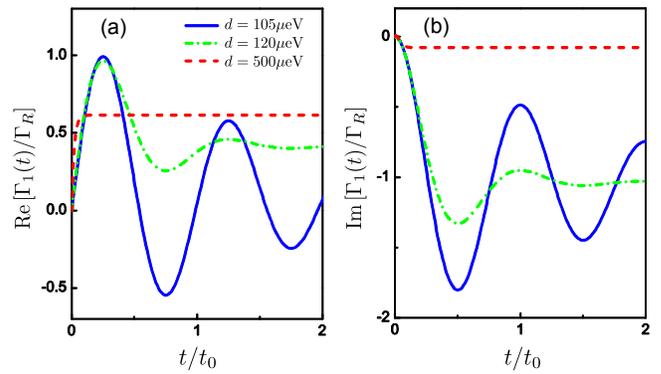} \caption{(Color
online) (a)~Real and (b)~imaginary parts of the time-dependent
coefficient $\Gamma_{1}(t)$ in the non-Markovian master equation of
a single quantum dot system, where $\Gamma_{L}=\Gamma_{R}= 100\mu
eV$, $\omega_{0}=50\mu eV$, and $t_{0}=2\pi/\omega_{0}$. }
\label{fig1}
\end{figure}

Figure~\ref{fig1} presents both real and imaginary parts of the
time-dependent coefficients $\Gamma_{1}(t)$ in
Eq.~(\ref{dot-master}). For simplicity, we consider the large bias
regime (i.e., $\mu_{L}\gg\omega_{0}\gg\mu_{R}$) and a zero
temperature for fermionic environments. The tunneling rates are
chosen in the symmetric case of $\Gamma_{L}=\Gamma_{R}=\Gamma$, so
that $\Gamma_{2}(t)=- \Gamma_{1}(t)$. Moreover, the noise is modeled
as the Ornstein-Uhlenbeck process, and the related correlation
functions are then given by
\begin{eqnarray}
&&\alpha_{L2}(t,s)= \alpha_{R1}(t,s)=
\frac{\Gamma d}{2}e^{-d\lvert t-s\rvert},\nonumber\\
&&\alpha_{L1}(t,s)=\alpha_{R2}(t,s)= 0,
\end{eqnarray}
where $1/d$ characterizes the memory time of
each environment. In the Markovian limit with $d\rightarrow\infty$,
$\alpha_{L2}(t,s)=\alpha_{R1}(t,s)\rightarrow\Gamma\delta(t-s)$, and
$\Gamma_{1}(t)$ becomes time-dependent. Indeed, Fig.~\ref{fig1}
shows that $\Gamma_{1}$ oscillates with time $t$, but it quickly
decays to a constant for a large value of $d$.

\subsection{Double quantum dot (DQD)}

Suppose that the DQD is in the
strong Coulomb blockade regime, so that at most one electron is allowed in
each dot. The Hamiltonian of the DQD can be written as
\begin{equation}
H_{\rm sys}= \omega_{1}c_{1}^\dagger c_{1}+ \omega_{2}c_{2}^\dagger
c_{2}+ \Omega_{0}(c_{2}^\dagger c_{1}+ c_{1}^\dagger c_{2}),
\end{equation}
where $\Omega_{0}$ denotes the inderdot coupling. For $H_{\rm int}$
in Eq.~(\ref{int}), $c_{L}= c_{1}$, and $c_{R}= c_{2}$. The exact
non-Markovian master equation is given by
\begin{align}
\frac{\partial\rho_{t}}{\partial t} =&-i[H_{\rm sys},\rho_{t}]
+\left\{\left(\Gamma_{L1}(t)[c_{1},\rho_{t}c_{1}^\dagger]
+\Gamma_{L2}(t)[c_{1},c_{1}^\dagger\rho_{t}]\right.\right.\nonumber\\
&+\Gamma_{L3}(t)[c_{1},\rho_{t}c_{2}^\dagger]
+\Gamma_{L4}(t)[c_{1},c_{2}^\dagger\rho_{t}]\nonumber\\
&+\Gamma_{R1}(t)[c_{2},\rho_{t}c_{1}^\dagger]
+\Gamma_{R2}(t)[c_{2},c_{1}^\dagger\rho_{t}]\nonumber\\
&+\left.\left.\Gamma_{R3}(t)[c_{2},\rho_{t}c_{2}^\dagger]
+\Gamma_{R4}(t)[c_{2},c_{2}^\dagger\rho_{t}]\right)+ {\rm H.c.}\right\}, \label{DQD-master}
\end{align}
with time-dependent coefficients
\begin{equation}
\Gamma_{\lambda j}(t)= \int_{0}^t[\alpha_{\lambda 1}(s-t)A_{\lambda
j}(t,s)-\alpha_{\lambda 2}(t-s)B_{\lambda j}(t,s)]ds,
\end{equation}
where $A_{\lambda j}(t,s)$ and $B_{\lambda j}(t,s)$ satisfy a set of
integro-differential equations (see Appendix B), with the final
conditions: $A_{L1}(t,t)=A_{R3}(t,t)=B_{L2}(t,t)=B_{R4}(t,t)=1$, and
$A_{\lambda j}(t,t)=0$, $B_{\lambda j}(t,t)=0$ for other $\lambda$
and $j$. A similar non-Markovian master equation was also obtained
using the Feynman-Vernon influence functional path-integral
method~\cite{Zhang1}.

In the Markovian limit, there are
\begin{eqnarray}
\Gamma_{L1}(t)\!&\!\rightarrow\!&\! \frac{1}{2}
[1-\bar{n}_{L}(\omega_{1})]\Gamma_{L},~~ \Gamma_{R1}\rightarrow
0; \nonumber\\
\Gamma_{L2}(t)\!&\!\rightarrow\!&\! -\frac{1}{2}\bar{n}_{L}(\omega_{1})
\Gamma_{L},~~ \Gamma_{R2}\rightarrow 0; \nonumber\\
\Gamma_{L3}(t)\!&\!\rightarrow\!&\! 0,~~ \Gamma_{R3}\rightarrow \frac{1}{2}[1-
\bar{n}_{R}(\omega_{2})]\Gamma_{R}; \\
\Gamma_{L4}(t)\!&\!\rightarrow\!&\! 0,~~ \Gamma_{R4}\rightarrow
-\frac{1}{2}\bar{n}_{R}(\omega_{2})]\Gamma_{R}. \nonumber
\end{eqnarray}
We also consider
the zero-temperature case with
$\bar{n}_{\lambda}(\omega_n)\rightarrow \theta(\mu_{\lambda}-
\omega_n)$, and the two single-dot levels of the DQD all lie within
the energy window $\mu_{L}>\omega_n>\mu_{R}$, where $n=1,2$. We use
$|l\rangle$, $l=0,1,2$ and 3, to denote the states with both dots
empty, the left dot occupied, the right dot occupied, and both dots
occupied, respectively. From Eq.~(\ref{DQD-master}), it follows that
the master equations for density matrix elements are reduced to
\begin{align}
&\dot{\rho}_{00}=-\Gamma_{L}\rho_{00}+ \Gamma_{R}\rho_{22},~~~\dot{\rho}_{33}=\Gamma_{L}\rho_{22}-\Gamma_{R}\rho_{33},\nonumber\\
&\dot{\rho}_{11}=\Gamma_{L}\rho_{00}+\Gamma_{R}\rho_{33}+i\Omega_{0}(\rho_{12}- \rho_{21}),\nonumber\\
&\dot{\rho}_{22}=-(\Gamma_{L}+\Gamma_{R})\rho_{22}- i\Omega_{0}(\rho_{12}- \rho_{21}),\\
&\dot{\rho}_{12}=-i(\omega_{1}- \omega_{2})\rho_{12}+
i\Omega_{0}(\rho_{11}- \rho_{22})-
\frac{\Gamma_L+\Gamma_R}{2}\rho_{12},\nonumber
\end{align}
which are identical to the rate equations obtained in
\cite{Gurvitz}. For a DQD, both intradot and interdot Coulomb
repulsions can play an important role in the Coulomb-blockade effect
(see, e.g., \cite{jqyou2}). Thus, if both intradot and interdot
Coulomb repulsions are so strong that only one electron is allowed
in the whole DQD, the master equations for density matrix elements
are reduced to
\begin{align}
&\dot{\rho}_{00}=-\Gamma_{L}\rho_{00}+ \Gamma_{R}\rho_{22}\nonumber\\
&\dot{\rho}_{11}=\Gamma_{L}\rho_{00}+ i\Omega_{0}(\rho_{12}- \rho_{21})\nonumber\\
&\dot{\rho}_{22}=-\Gamma_{R}\rho_{22}- i\Omega_{0}(\rho_{12}- \rho_{21})\\
&\dot{\rho}_{12}= -i(\omega_{1}-\omega_{2})\rho_{12}+
i\Omega_{0}(\rho_{11}- \rho_{22})-
\frac{\Gamma_{R}}{2}\rho_{12},\nonumber
\end{align}
which are exactly the rate equations obtained in \cite{Stoof}.

\section{Conclusion}

We have developed an NMQSD method to study
the dynamics of an open quantum system in fermionic baths. We not
only obtain the NMQSD equation for quantum states of the system, but
also derive the non-Markovian master equation. This non-Markovian
approach is formulated in a non-perturbative manner and it applies
even for a strong coupling between the system and the fermionic
baths. Moreover, as useful examples, we have applied this NMQSD
method to single and double quantum-dot systems.

{\it Note added}: After finishing this work, we were aware of a
closely related work in~\cite{Yu3}, which also investigated an open
quantum system in fermionic baths by using a quantum state diffusion
approach.

\begin{acknowledgments}
This work was supported by the National Basic Research Program of
China Grant No. 2009CB929302 and the National Natural Science
Foundation of China Grant No. 91121015.
\end{acknowledgments}
%

\appendix
\section{Non-Markovian master equation for a single quantum dot sytem}
In this appendix, we show how to obtain the exact non-Markovian master equation for a single quantum dot system from the general non-Markovian master equation~(\ref{g-master}) in the main text. In order to get a closed
convolutionless master equation, we need to know the explicit forms
of the operator functions
\begin{align}
& \mathcal{Q}_{1}(t,s)\equiv \mathcal{M}\{O_{\lambda 1}(t,s,z^*,w^*)P_{t}\},\nonumber\\
& \mathcal{Q}_{2}(t,s)\equiv \mathcal{M}\{O_{\lambda 2}(t,s,z^*,w^*)P_{t}\},
\end{align}
and their hermitian conjugates. Similar to the Heisenberg equation
approach for quantum state diffusion in a bosonic environment
\cite{Strunz2,Yu2}, we start to obtain the evolution equations for
\begin{align}
&\mathcal{C}_{1}(s)= \langle zw\rvert\mathcal{U}_{t}c(s)\lvert 0\rangle= O_{\lambda 1}(t,s,z^*,w^*)G_{t}(z^*,w^*),\nonumber\\
\label{c02}&\mathcal{C}_{2}(s)= \langle zw\rvert\mathcal{U}_{t}c^\dagger(s)\lvert 0\rangle= O_{\lambda 2}(t,s,z^*,w^*)G_{t}(z^*,w^*),
\end{align}
where $\mathcal{U}_{t}$ is the time evolution operator for the total
system: $\lvert\Psi_{t}\rangle=
\mathcal{U}_{t}\lvert\Psi_{0}\rangle$. The stochastic propagator of
the open quantum system can be expressed using the matrix element of
the time evolution operator: $G_{t}(z^*,w^*)= \langle
zw\rvert\mathcal{U}_{t}\lvert 0\rangle$. Now we take the operators
$\mathcal{C}_{i}$ ($i=1,2$) as functions of $s$ and treat $t$ only
as a parameter. With the Heisenberg equation of motion for the
fermionic operator $c(s)$ of the single quantum dot:
\begin{equation}
\frac{\partial}{\partial s}c(s)=
-i\mathcal{U}_{s}^{-1}[c,\mathcal{H}(s)]\mathcal{U}_{s},
\end{equation}
we have
\begin{align}
\frac{\partial}{\partial s}\mathcal{C}_{1}(s)=&
-i\omega_{0}\mathcal{C}_{1}(s)\nonumber\\
&-i\sum_{\lambda k}\sqrt{1-
\bar{n}_{\lambda k}}g_{\lambda k}e^{-i\omega_{\lambda k}s}\langle
zw\rvert \mathcal{U}_{t}d_{\lambda k}(s)\lvert 0\rangle\nonumber\\
&+i\sum_{\lambda k}\sqrt{\bar{n}_{\lambda k}}g_{\lambda
k}e^{-i\omega_{\lambda k}s} \langle zw\rvert
\mathcal{U}_{t}e_{\lambda k}^\dagger(s)\lvert 0\rangle.
\label{c1}
\end{align}
By integrating the Heisenberg equations of motion for both
$d_{\lambda k}$ and $e_{\lambda k}^\dagger$, we can get
\begin{align}
& d_{\lambda k}(s)= d_{\lambda k}- i\sqrt{1- \bar{n}_{\lambda k}}
g_{\lambda k}^*\int_{0}^s e^{i\omega_{\lambda k}s'}c(s')ds',\nonumber\\
\label{e}& e_{\lambda k}^\dagger(t)= e_{\lambda k}^\dagger(s)+ i\sqrt{\bar{n}_{\lambda k}}
g_{\lambda k}^*\int_{s}^t e^{i\omega_{\lambda k}s'}c(s')ds'.
\end{align}
Substituting Eq.~(\ref{e}) into Eq.(\ref{c1}), we derive the evolution
equation for $\mathcal{C}_{1}(s)$ as
\begin{align}
\frac{\partial}{\partial s}\mathcal{C}_{1}(s)=&\!-i\omega_{0}\mathcal{C}_{1}(s)\nonumber\\
&- \int_{0}^s[\alpha_{L1}(s-s')
+ \alpha_{R1}(s-s')]\mathcal{C}_{1}(s')ds'\nonumber\\
&+ \int_{s}^t[\alpha_{L2}(s'-s)
+ \alpha_{R2}(s'-s)]\mathcal{C}_{1}(s')ds'\nonumber\\
&- [w_{L}^*(s)+ w_{R}^*(s)]\langle zw\rvert\mathcal{U}(t)\lvert 0\rangle.
\end{align}
%
%
%
%
%
In the above calculations, we have used the relations $d_{\lambda k}\lvert 0\rangle= 0$,
and $\langle zw\rvert\mathcal{U}_{t}e_{\lambda k}^\dagger(t)\lvert 0\rangle
= w_{\lambda k}^*\langle zw\rvert \mathcal{U}_{t}\lvert 0\rangle$.
Similarly, we can also derive the evolution equation for $\mathcal{C}_{2}(s)$ as
\begin{align}
\frac{\partial}{\partial s}\mathcal{C}_{2}(s)=&\ i\omega_{0}\mathcal{C}_{2}(s)\nonumber\\
&+\int_{s}^t[\alpha_{L1}(s'-s)+ \alpha_{R1}(s'-s)]\mathcal{C}_{2}(s')ds'\nonumber\\
&-\int_{0}^s[\alpha_{L2}(s-s')+ \alpha_{R2}(s-s')]\mathcal{C}_{2}(s')ds'\nonumber\\
&- [z_{L}^*(s)+ z_{R}^*(s)]\langle zw\rvert\mathcal{U}(t)\lvert 0\rangle.
\end{align}

From Eq.~(\ref{c02}), we can thus
obtain the evolution equation for the $O$-operators
$O_{\lambda 1}(t,s,z^*,w^*)$ and
$O_{\lambda 2}(t,s,z^*,w^*)$ :
\begin{align}
&\frac{\partial}{\partial s}O_{\lambda 1}(t,s,z^*,w^*)= -i\omega_{0} O_{\lambda 1}(t,s,z^*,w^*)\nonumber\\
&- \int_{0}^s[\alpha_{L1}(s-s')+ \alpha_{R1}(s-s')]O_{\lambda 1}(t,s',z^*,w^*)ds'\nonumber\\
\label{O uncoupled 1}&+ \int_{s}^t[\alpha_{L2}(s'-s)+ \alpha_{R2}(s'-s)]O_{\lambda 1}(t,s',z^*,w^*)ds'\nonumber\\
&- [w_{L}^*(s)+ w_{R}^*(s)],
\end{align}
and
\begin{align}
&\frac{\partial}{\partial s}O_{\lambda 2}(t,s,z^*,w^*)= i\omega_{0} O_{\lambda 2}(t,s,z^*,w^*)\nonumber\\
&+ \int_{s}^t[\alpha_{L1}(s'-s)+ \alpha_{R1}(s'-s)]O_{\lambda 2}(t,s',z^*,w^*)ds'\nonumber\\
\label{O uncoupled 2}&- \int_{0}^s[\alpha_{L2}(s-s')+ \alpha_{R2}(s-s')]O_{\lambda 2}
(t,s',z^*,w^*)ds'\nonumber\\
&- [z_{L}^*(s)+ z_{R}^*(s)],
\end{align}
with final conditions at $s=t$:
\begin{equation}
O_{\lambda1}(t,t,z^*,w^*)= c,\quad O_{\lambda2}(t,t,z^*,w^*)= c^\dagger.
\end{equation}
By taking the mean $\mathcal{M}\{\cdots P_{t}\}$ on
Eqs.~(\ref{O uncoupled 1}) and~(\ref{O uncoupled 2}), and using the relations
\begin{align}
&\mathcal{M}\{w_{\lambda}^*(s)P_{t}\}
= -\mathcal{M}\{P_{t}\bar{O}_{\lambda 2}^\dagger(t,-z,-w)\}\nonumber\\
&\mathcal{M}\{z_{\lambda}^*(s)P_{t}\}
= -\mathcal{M}\{P_{t}\bar{O}_{\lambda 1}^\dagger(t,-z,-w)\},
\end{align}
we derive the evolution equations for the operator functions
$\mathcal{Q}_{n}(s)$ $(n=1,2)$ as
\begin{align}
&\frac{\partial}{\partial s}\mathcal{Q}_{1}(t,s)= -i\omega_{0} \mathcal{Q}_{1}(t,s)\nonumber\\
&- \int_{0}^s[\alpha_{L1}(s-s')+ \alpha_{R1}(s-s')]\mathcal{Q}_{1}(t,s')ds'\nonumber\\
&+ \int_{s}^t[\alpha_{L2}(s'-s)+ \alpha_{R2}(s'-s)]\mathcal{Q}_{1}(t,s')ds'\nonumber\\
\label{Q1}&\ + \int_{0}^t[\alpha_{L2}^*(s-s')
 + \alpha_{R2}^*(s-s')]\mathcal{Q}_{2}^\dagger(t,s')ds',
\end{align}
and
\begin{align}
&\frac{\partial}{\partial s}\mathcal{Q}_{2}(t,s)=i\omega_{0} \mathcal{Q}_{2}(t,s)\nonumber\\
&+ \int_{s}^t[\alpha_{L1}(s'-s)+ \alpha_{R1}(s'-s)]\mathcal{Q}_{2}(t,s')ds'\nonumber\\
&- \int_{0}^s[\alpha_{L2}(s-s')+ \alpha_{R2}(s-s')]\mathcal{R}_{2}(t,s')ds'\nonumber\\
\label{Q2}&+ \int_{0}^t[\alpha_{L1}^*(s-s')+ \alpha_{R1}^*(s-s')]\mathcal{Q}_{1}^\dagger(t,s')ds',
\end{align}
with final conditions at $s=t$: $\mathcal{Q}_{1}(t,t)= c\rho_{t}$, and $\mathcal{Q}_{2}(t,t)= c^\dagger\rho_{t}$.

According to Eqs.~(\ref{Q1}) and (\ref{Q2}) and the final conditions of
the operator functions $\mathcal{Q}_{n}(t,s)$, $n=1,2$, it can be
seen that $\mathcal{Q}_{1}(t,s)$ and $\mathcal{Q}_{2}(t,s)$ should
have the forms
\begin{align}
& \mathcal{Q}_{1}(t,s)= A_{1}^*(t,s)c\rho_{t}+ A_{2}^*(t,s)\rho_{t}c,\nonumber\\
\label{qb}& \mathcal{Q}_{2}(t,s)= B_{1}(t,s)c^\dagger\rho_{t}+ B_{2}(t,s)\rho_{t}c^\dagger.
\end{align}
Substituting $\mathcal{Q}_{1}(t,s)$ and $\mathcal{Q}_{2}(t,s)$ and their hermitian conjugates into Eq.~(\ref{g-master}),
we finally obtain the non-Markovian master equation for a single quantum dot system
\begin{align}
\frac{\partial}{\partial t}\rho_{t}=&\ -i\omega_{0}[c^\dagger c,\rho_{t}]
+ \Gamma_{1}(t)[c,\rho_{t}c^\dagger]+ \Gamma_{2}(t)[c,c^\dagger\rho_{t}]\nonumber\\
&\ + \Gamma_{1}^*(t)[c\rho_{t},c^\dagger]+ \Gamma_{2}^*(t)[\rho_{t}c,c^\dagger],
\end{align}
which is just Eq.~(\ref{dot-master}) in the main text. Here the time-dependent
coefficients $\Gamma_{j}(t)$, $j=1$ and 2, are given in Eq.~(\ref{coefficient a}).
%
%
%
%
%
%
%

\section{Non-Markovian master equation for a double quantum dot system}
As in Appendix~A, by the Heisenberg equation approach, it can be
derived for the double quantum dot system that the $O$-operators
satisfy the following integro-differential equations:
\begin{align}
&\frac{\partial}{\partial s}O_{L1}(t,s,z^*,w^*)= -i\omega_{1} O_{L1}(t,s,z^*,w^*)\nonumber\\
&~~~- i\Omega_{0}O_{R1}(t,s,z^*,w^*)\nonumber\\
&~~~- \int_{0}^s\alpha_{L1}(s-s')O_{L1}(t,s',z^*,w^*)ds'\nonumber\\
&~~~+\int_{s}^t\alpha_{L2}(s'-s)O_{L1}(t,s',z^*,w^*)ds'
-w_{L}^*(s),\nonumber\\
%
&\frac{\partial}{\partial s}O_{R1}(t,s,z^*,w^*)= -i\omega_{2} O_{R1}(t,s,z^*,w^*)\nonumber\\
&~~~- i\Omega_{0}O_{L1}(t,s,z^*,w^*)\nonumber\\
&~~~- \int_{0}^s\alpha_{R1}(s-s')O_{R1}(t,s',z^*,w^*)ds'\nonumber\\
\label{OL2}&~~~+\int_{s}^t\alpha_{R2}(s'-s)O_{R1}(t,s',z^*,w^*)ds'
-w_{R}^*(s),
\end{align}
and
\begin{align}
&\frac{\partial}{\partial s}O_{L2}(t,s,z^*,w^*)= i\omega_{1}
O_{L2}(t,s,z^*,w^*)\nonumber\\
&~~~+ i\Omega_{0}O_{R2}(t,s,z^*,w^*)\nonumber\\
&~~~+ \int_{s}^t\alpha_{L1}(s'-s)O_{L2}(t,s',z^*,w^*)ds'\nonumber\\
&~~~-\int_{0}^s\alpha_{L2}(s-s')O_{L2}(t,s,z^*,w^*)ds'-z_{L}^*(s),\nonumber\\
&\frac{\partial}{\partial s}O_{R2}(t,s,z^*,w^*)= i\omega_{2}
O_{R2}(t,s,z^*,w^*)\nonumber\\
&~~~+ i\Omega_{0}O_{L2}(t,s,z^*,w^*)\nonumber\\
&~~~+ \int_{s}^t\alpha_{R1}(s'-s)O_{R2}(t,s',z^*,w^*)ds'\nonumber\\
\label{OR2}&~~~-\int_{0}^s\alpha_{R2}(s-s')O_{R2}(t,s,z^*,w^*)ds'-z_{R}^*(s),
\end{align}
with final conditions at $s=t$: $O_{L1}(t,t,z^*,w^*)= c_{1}$,
$O_{R1}(t,t,z^*,w^*)= c_{2}$, $O_{L2}(t,t,z^*,w^*)= c_{1}^\dagger$,
and $O_{R2}(t,t,z^*,w^*)= c_{2}^\dagger$.

Define $\mathcal{Q}_{\lambda n}(t,s)\equiv \mathcal{M}\{O_{\lambda
n}(t,s,z^*,w^*)P_{t}\}$, where $\lambda=L,R$, and $n=1,2$.
 From Eqs.~(\ref{OL2}) and (\ref{OR2}), the
evolution equations for the operator functions $\mathcal{Q}_{\lambda
n}(t,s)$ can be derived as
\begin{align}
\frac{\partial}{\partial s}\mathcal{Q}_{L1}(t,s)=& -i\omega_{1}
\mathcal{Q}_{L1}(t,s)- i\Omega_{0}\mathcal{Q}_{R1}(t,s)\nonumber\\
&- \int_{0}^s\alpha_{L1}(s-s')\mathcal{Q}_{L1}(t,s')ds'\nonumber\\
&+ \int_{s}^t\alpha_{L2}(s'-s)\mathcal{Q}_{L1}(t,s')ds'\nonumber\\
&+
\int_{0}^t\alpha_{L2}^*(s-s')\mathcal{Q}_{L2}^\dagger(t,s')ds',\nonumber\\
%
\frac{\partial}{\partial s}\mathcal{Q}_{R1}(t,s)=& -i\omega_{2}
\mathcal{Q}_{R1}(t,s)- i\Omega_{0}\mathcal{Q}_{L1}(t,s)\nonumber\\
&- \int_{0}^s\alpha_{R1}(s-s')\mathcal{Q}_{R1}(t,s')ds'\nonumber\\
&+ \int_{s}^t\alpha_{R2}(s'-s)\mathcal{Q}_{R1}(t,s')ds'\nonumber\\
\label{ql2}&+\int_{0}^t\alpha_{R2}^*(s-s')\mathcal{Q}_{R2}^\dagger(t,s')ds',
\end{align}
and
\begin{align}
\frac{\partial}{\partial s}\mathcal{Q}_{L2}(t,s)=&\; i\omega_{1}
\mathcal{Q}_{L2}(t,s)+ i\Omega_{0}\mathcal{Q}_{R2}(t,s)\nonumber\\
&+\int_{s}^t\alpha_{L1}(s'-s)\mathcal{Q}_{L2}(t,s')ds'\nonumber\\
&- \int_{0}^s\alpha_{L2}(s-s')\mathcal{Q}_{L2}(t,s')ds'\nonumber\\
&+ \int_{0}^t\alpha_{L1}^*(s-s')\mathcal{Q}_{L1}^\dagger(t,s')ds',\nonumber\\
%
\frac{\partial}{\partial s}\mathcal{Q}_{R2}(t,s)=&\; i\omega_{2}
\mathcal{Q}_{R2}(t,s)+ i\Omega_{0}\mathcal{Q}_{L2}(t,s)\nonumber\\
&+\int_{s}^t\alpha_{R1}(s'-s)\mathcal{Q}_{R2}(t,s')ds'\nonumber\\
&- \int_{0}^s\alpha_{R2}(s-s')\mathcal{Q}_{R2}(t,s')ds'\nonumber\\
\label{qr2}&+ \int_{0}^t\alpha_{R1}^*(s-s')\mathcal{Q}_{R1}^\dagger(t,s')ds'.
\end{align}
Correspondingly, the final conditions of $\mathcal{Q}_{\lambda
n}(t,s)$ at $s=t$ are given by $\mathcal{Q}_{L1}(t,t)=
c_{1}\rho_{t}$, $\mathcal{Q}_{L2}(t,t)= c_{1}^\dagger\rho_{t}$,
$\mathcal{Q}_{R1}(t,t)= c_{2}\rho_{t}$, and $\mathcal{Q}_{R2}(t,t)=
c_{2}^\dagger\rho_{t}$.

According to Eqs.~(\ref{ql2}) and (\ref{qr2}) and the final conditions
of the operator functions $\mathcal{Q}_{\lambda n}(t,s)$,
$\mathcal{Q}_{\lambda n}(t,s)$ should take the forms
\begin{align}
\mathcal{Q}_{\lambda 1}(t,s)=& A_{\lambda 1}^*(t,s)c_{1}\rho_{t} +
A_{\lambda 2}^*(t,s)\rho_{t}c_{1}+ A_{\lambda 3}^*(t,s)c_{2}\rho_{t}\nonumber\\
&+ A_{\lambda4}^*(t,s)\rho_{t}c_{2},\nonumber\\
\label{Q_2}\mathcal{Q}_{\lambda 2}(t,s)=& B_{\lambda
1}(t,s)\rho_{t}c_{1}^\dagger+B_{\lambda 2}(t,s)c_{1}^\dagger\rho_{t}
+B_{\lambda 3}(t,s)\rho_{t}c_{2}^\dagger\nonumber\\
&+ B_{\lambda4}(t,s)c_{2}^\dagger\rho_{t},
\end{align}
where $\lambda=L,R$, and the final conditions of $A_{\lambda
j}(t,s)$ and $B_{\lambda j}(t,s)$ are given as follows:
$A_{L1}(t,t)=A_{R3}(t,t)=B_{L2}(t,t)=B_{R4}(t,t)=1$, and $A_{\lambda
j}(t,t)=0$, $B_{\lambda j}(t,t)=0$ for other $\lambda$ and $j$. Here
$A_{\lambda j}(t,s)$ and $B_{\lambda j}(t,s)$ satisfy the following
integro-differential equations:
\begin{align}
&\frac{\partial}{\partial s}A_{Lj}(t,s)- i\omega_{1}A_{Lj}(t,s)
- i\Omega_{0}A_{Rj}(t,s)\nonumber\\
&\quad+ \int_{0}^s\beta_{L}(s-s')A_{Lj}(t,s')ds'
= U_{Lj}(t,s),\nonumber\\
&\frac{\partial}{\partial s}A_{Rj}(t,s)- i\omega_{2}A_{Rj}(t,s)
- i\Omega_{0}A_{Lj}(t,s)\nonumber\\
&\quad+ \int_{0}^s\beta_{R}(s-s')A_{Rj}(t,s')ds'
= U_{Rj}(t,s),
\end{align}
and
\begin{align}
&\frac{\partial}{\partial s}B_{Lj}(t,s)- i\omega_{1}B_{Lj}(t,s)
-i\Omega_{0}B_{Rj}(t,s)\nonumber\\
&\quad+ \int_{0}^s\beta_{L}(s'-s)B_{Lj}(t,s')ds'
= V_{Lj}(t,s),\nonumber\\
&\frac{\partial}{\partial s}B_{Rj}(t,s)- i\omega_{2}B_{Rj}(t,s)
-i\Omega_{0}B_{Lj}(t,s)\nonumber\\
&\quad+ \int_{0}^s\beta_{R}(s'-s)B_{Rj}(t,s')ds' =
V_{Rj}(t,s),
\end{align}
with
\begin{align}
& U_{L j}(t,s)=\int_{0}^t\alpha_{L 2}(s-s')h_{L 1}(t,s')ds',\nonumber\\
&U_{R j}(t,s)=\int_{0}^t\alpha_{R 2}(s-s')h_{R 2}(t,s')ds',
\quad {\rm for}~j=1,2, \nonumber\\
%
&U_{L j}(t,s)=\int_{0}^t\alpha_{L 2}(s-s')h_{L 2}(t,s')ds',\nonumber\\
\label{ul}&U_{R j}(t,s)=\int_{0}^t\alpha_{R 2}(s-s')h_{R 1}(t,s')ds',
\quad {\rm for}~j=3,4,
\end{align}
and
\begin{align}
& V_{L j}(t,s)=\int_{0}^t\alpha_{L 1}(s-s')h_{L 1}(t,s')ds',\nonumber\\
&V_{R j}(t,s)=\int_{0}^t\alpha_{L 1}(s-s')h_{R 2}(t,s')ds',
\quad {\rm for}~j=1,2, \nonumber\\
%
\label{vl}&V_{L j}(t,s)=\int_{0}^t\alpha_{L 1}(s-s')h_{L 2}(t,s')ds',\nonumber\\
&V_{R j}(t,s)=\int_{0}^t\alpha_{R 1}(s-s')h_{R 1}(t,s')ds', \quad
{\rm for}~j=3,4.
\end{align}
In Eqs.~(\ref{ul}) and (\ref{vl}), $h_{Ln}(t,s)$ and $h_{Rn}(t,s)$, with
$n=1,2$, are solutions of the two coupled integro-differential
equations:
\begin{align}
\frac{\partial}{\partial s}h_{Ln}(t,s)=&\ i\omega_{1}h_{Ln}(t,s)
+ i\Omega_{0}h_{Rn}(t,s)\nonumber\\
&+ \int_{s}^t\beta_L(s-s')h_{Ln}(t,s')ds',\nonumber\\
\frac{\partial}{\partial s}h_{Rn}(t,s)=&\ i\omega_{2}h_{Rn}(t,s)+
i\Omega_{0}h_{Ln}(t,s)\nonumber\\
&+\int_{s}^t\beta_R(s-s')h_{Rn}(t,s')ds',
\end{align}
with final conditions at $s=t$: $h_{L1}(t,t)= h_{R1}(t,t)= 1$, and
$h_{L2}(t,t)= h_{R2}(t,t)= 0$. Here $\beta_{\lambda}(s-s')$ is
defined by
\begin{equation}
\beta_{\lambda}(s-s')\equiv \alpha_{\lambda 1}(s'-s)+
\alpha_{\lambda 2}(s-s').
\end{equation}

Using the obtained $\mathcal{Q}_{\lambda n}(t,s)$ in Eqs.~(\ref{Q_2}), one
can calculate
\begin{align}
&\mathcal{M}\{\bar{O}_{\lambda n}(t,z^*,w^*)P_{t}\}=\int_0^t ds
\alpha_{\lambda n}(t-s)\mathcal{Q}_{\lambda n}(t,s),\nonumber\\
&\mathcal{M}\{P_t\bar{O}_{\lambda n}^{\dagger}(t,-z,-w)\}=\int_0^t
ds \alpha_{\lambda n}(t-s)\mathcal{Q}_{\lambda
n}^{\dagger}(t,s).
\label{B12}
\end{align}
Substituting Eq.~(\ref{B12})
into
Eq.~(\ref{g-master}), we then obtain the exact master equation in Eq.~(\ref{DQD-master}) for
the double-quantum-dot system.


\end{document}